# Design, Fabrication, and Testing of a Liquid Cooling Platform for High Power 3D-ICs

Chandrasekhar Mandalapu
*Department of Electrical Engineering*
Northern Illinois University
DeKalb, IL 60115 USA
Chandu.mandalapu@gmail.com

Ibrahim M Abdel-Motaleb[*]
*Department of Electrical Engineering*
Northern Illinois University
DeKalb, IL 60115 USA
ibrahim@niu.edu

Sangki Hong
Nhanced-Semiconductor, Inc.
Naperville, IL 60563 USA
shong@nhanced-semi.com

Robert Patti
Nhanced-Semiconductor, Inc.
Naperville, IL 60563 USA
rpatti@nhanced-semi.com

*Abstract*—3D integrated circuit (3D-IC) technology gained acceptance due to the ability to achieve extremely high level of integration, where hundreds of ICs are stacked vertically. Such level of integration can result in local power dissipation of more than 50 kW/cm$^2$. This will lead to instant evaporation of the IC, unless an effective cooling technique is employed. Liquid cooling may be one of the most effective techniques for this task. To investigate the effectiveness of this technique, we designed, built and tested a testing platform. The platform includes a testing chip and a cooling module. The test chip contains heaters to provide the power and sensors to measure the local temperature. The study shows that the proposed cooling modules can reduce the temperature for a 420W/inch square circuits to a normal operating range of ICs of 39-50 °C, using 2 phase R22 liquid coolant.

## I. Introduction

3D integrated circuit (3D-IC) chips consist of multiple thinned-active 2D integrated circuits that are stacked, bonded, and electrically connected with vertical vias formed through silicon or oxide layers [1]. Because of the stacking of the 2D IC, 3D ICs can increase the integration density several folds. The main drawback of 3D ICs is the extremely high local power dissipation that can reach 1000 kW/cm$^2$ [2]. Such extreme power dissipation can push the temperature of some hot spots on the IC surface to serval 1000's of degrees in a fraction of a seconds [3]. This will defiantly result in an instant incineration of the IC. To avoid this fate, effective cooling technique capable of reducing hot spot temperatures to acceptable operating levels.

IC cooling can be passive or active. Passive cooling such as thermal conduction (e.g. metal lines), natural convection (e.g. finned heat sinks), or radiation (e.g. coating paints), are simple and inexpensive, but cannot handle such high thermal energy. Active cooling requires input power since it uses external components such as forced convection (using fans or nozzles), pumped loops (using heat exchangers or cold plates) and refrigerators (using Peltier/thermoelectric or vapor-compression). Active cooling can be an adaptive or non-adaptive system. Adaptive system incorporates a closed loop feedback of the temperature to create a temperature aware cooling system [4].

Current 3D IC cooling are mostly based on air cooling using heat-sink-fan or macrofluidics where the chip is immersed in liquid with MEMS macrochannels to remove heat [5]. Refrigeration systems such as thermoelectric devices or gas/liquid compression is also used [6]. The viability of each system depends on its ability to provide efficient heat management.

Microfluidics liquid cooling techniques are still in their early stages of development. Although there are many design concepts in the literature, they require practical implantation and rigorous testing. Testing on a real 3D IC is prohibitively costly. Using a platform or test bed is a viable alternative.

In this paper, we present the design, fabrication, and testing results of a test chip and a microchannel fluidic cooling block to test the cooling of 500W/cm$^2$ local hot spots in a 420W/inch-square IC.

## II. Design and Fabrication

The cooling block is bonded to the test chip as shown in Fig. 1. Using an 8-inch Si wafer with a SiO$_2$ layer grown on its surface, 32 test chips were built. Each test chip has an area of 25.6x26 mm$^2$. As shown in Fig. 2, each chip has thirty 10W and six 20W hot spot heaters spread across the chip. The 10-W heater area is 1x2 mm$^2$ while the 20W heater area is 2x2 mm$^2$. The test chip has also 42 temperature sensors, each with an area of 200x200 μm$^2$. The sensors are placed on the top of the 10W heaters and in between the heaters. Heaters and sensors shapes are serpentine. Heaters were fabricated using tungsten (W) and sensors using platinum (Pt). Cu is used for connection to the Al pads for both devices.

The physical parameters and dimensions for the heaters and sensor are shown in Table I. Sensors are connected to the Al pads with 50 μm width and 1μm thick copper wires. To reduce joule heating, heaters are connected to the Al pads with 250 μm width and 1μm thick copper wires. An SEM image of a cross section of the W-heater is shown in Fig. 3. The image shows the smoothness of the deposited metal and the uniformity of its thickness. Fig. 4 shows an SEM image of a cross section of the Pt-sensor metal. The image shows the metal stack composition and the corresponding thickness, including the top and bottom layer of SiO$_2$.

Since each chip has 36 heaters, we designed 36 cooling blocks, one for each heater, as shown in Fig. 5. The cooling block is composed of a 70-150 μm height-diamond coated microchannel, with pillars to create turbulences to increase the cooling efficiency. The block is made from bonding two substrates with 35-75 μm pillars, etched in the wafer. Fig. 6 shows one of the wafers with 35 μm diameter pillars etched using Deep Reactive Ion Etching (DRIE). Some are circular and others are rectangular. In some blocks, diamond was deposited on entire wafer, which is better thermal conductor and to enhance heat dissipation during liquid/gas flow. As it is very hard to do diamond polishing(CMP), a thin layer of $SiO_2$ is deposited on top of the diamond and CMP polished to improve the oxide roughness. Two patterned and CMP polished wafers are face-to-face bonded using plasma activated low temperature oxide-to-oxide bonding to create micro channels in cooling block, as shown in Fig.1

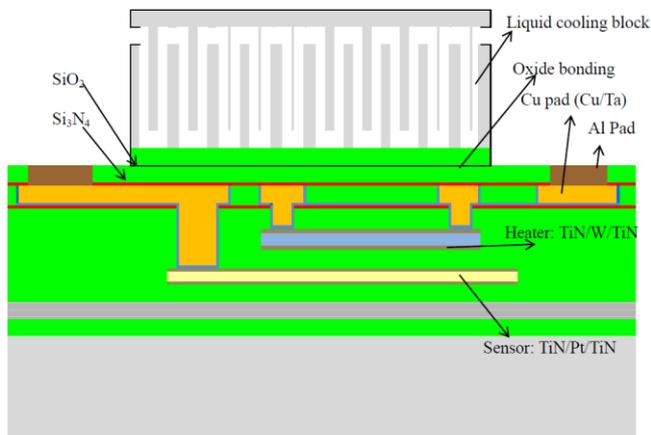

Fig. 1. Schematic diagram of the liquid cooling block attached on top of thermal test chip.

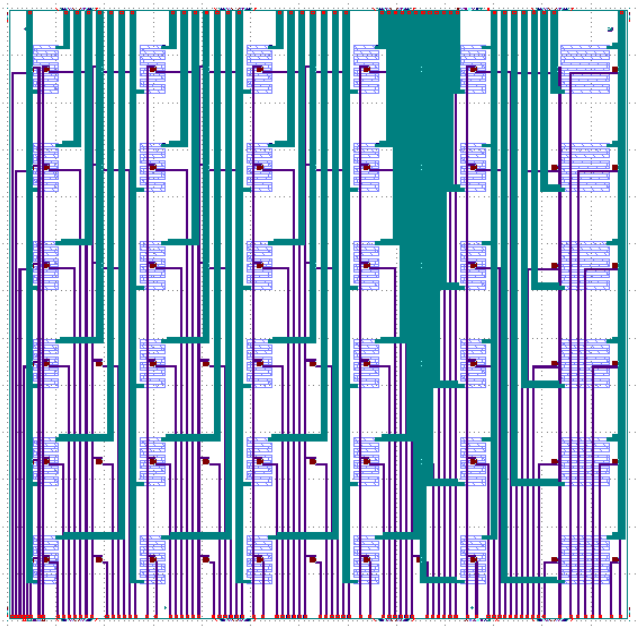

Fig. 2. Top view of the thermal test chip. Heaters are in blue, sensors are small brown squares, copper metal 1 in pink, copper metal 2 in green, and aluminum pads in red colors.

TABLE I. PHYSICAL PARAMETERS OF HEATERS AND SENSOR

| Parameter | Heater-10W | Heater 20W | Sensor |
|---|---|---|---|
| Area | 1x2 mm$^2$ | 2x2 mm$^2$ | 0.2x0.2 mm$^2$ |
| Resistance | 40 Ω | 80 Ω | 200 Ω |
| Line width | 120 μm | 120 μm | 10 μm |
| Spacing | 115 μm | 115 μm | 10 μm |
| Heater metal | TiN/W/TiN | TiN/W/TiN | Ti/TiN/Pt/TiN |
| Resistivity of stack | 9.52x10$^{-8}$ Ω.m | 9.52x10$^{-8}$ Ω.m | 11.93x10$^{-8}$ Ω.m |
| Total Length | 9920 μm | 18920 μm | 20900 μm |
| Thickness | 0.18 μm | 0.18 μm | 0.12 μm |

This work was supported by Tezzaron crop, Naperville, IL USA

## III. TESTING RESULTS

Current was applied to the heater terminals to generate specific heat flux. The corresponding temperature was then obtained from the change in the sensor resistance with heat.

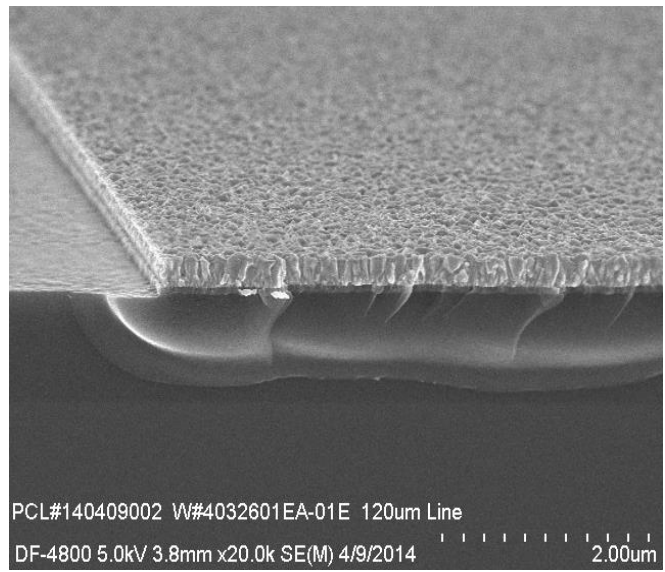

Fig. 3. SEM image of tungsten resistive heater

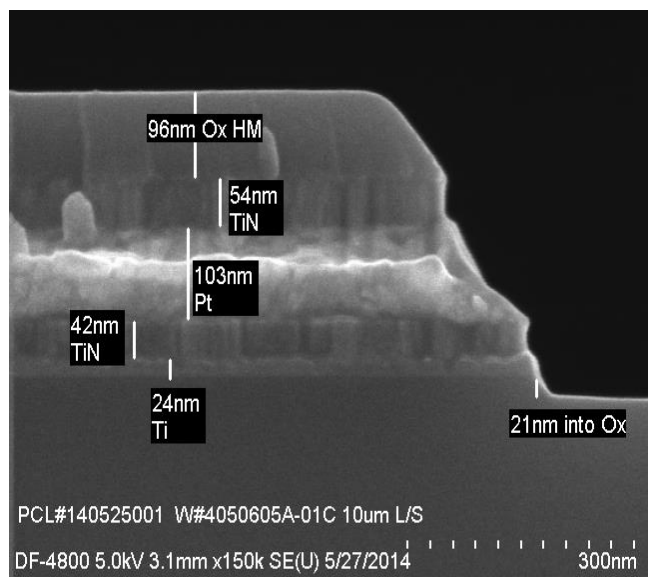

Fig. 4. SEM image of sensor metal stack. From top to bottom:96 nm oxide hard mask, 54 nm TiN/103 nm Pt/42 nm TiN/24 nm Ti. In addition, there are 96 nm oxide on the top and 21 nm at the bottom.

The testing results for a 10 W heater at the center of the wafer are shown in Table II. In this table, power ranging from 0 to 21.6 W were applied and the sensors resistance was measured and the corresponding temperatures were recorded from two sensors: left (between heaters) and center (on top of heater). When a current of 0.54 A is applied, the heater emitted 21.6 W. At this power the temperatures were found to be 124.23 ˚C for left sensor and 204.65 ˚C for center sensor. The high temperature of the center sensor is expected, since it is measuring the direct temperature of the hotspot. This behavior is true for all input powers.

This results seems to contradict the study reported in [3], where the temperature of the chip without cooling can reach up to 3250-4200 ˚C, for a hot spot power density of 500-1000 W/cm$^2$. This huge discrepancy is due to the fact that the study of [3] obtained the temperature for a heater built on 5x5 mm$^2$; while this study measured the temperature of a heater built on 8-inch diameter wafer. The wafer, in this case, acted as a huge heat sink, with an area of 50 inch-square, or 50 times the area of one chip. With powering all 36 heaters, the wafer area is 50 times the chip area. Assuming power dissipation is proportional to the heat sink area, an average temperature of 4200/50 = 84˚C and a hot spot temperature of 125-205 ˚C are reasonable.

R22 refrigerant was next used to test the cooling efficiency of the block. Individual chips are diced, packaged and tested. Therefore, the cooling effect of the wafer is eliminated. R22 liquid is applied to the inlet at 46 psi to ensure two phase operation (liquid & gas), since the boiling point is dropped to 25 ˚C. Increasing the inlet pressure, e.g. 100 psi, would increase the boiling point to 60 ˚C. This would, in turn, restrict the cooling process to the less efficient one phase process. The liquid is supplied from the cylinder and pushed into the cooling block. The resulting liquid and gas are collected from the outlet at atmospheric pressure.

Table III maps the measured temperature for each heater. As can be seen from the table, the temperatures of columns 1-5 (10 W-heaters) are 41-43 ˚C. The temperatures increase to 52 ˚C for the 20-W heater (column 6). This shows that the cooling system was able to reduce the temperature of a 420 W IC (microprocessor-like) to the normal operating temperature 25-60 ˚C for Si technology.

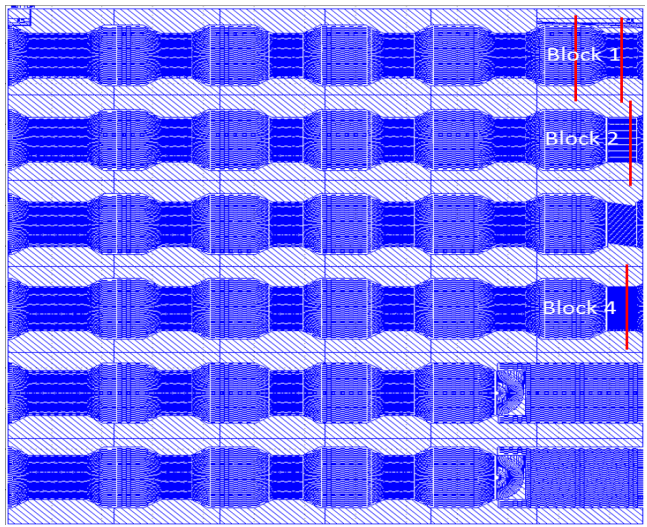

Fig. 5. Cooling block device depicting regions marked with red color of which SEM images are obtained

## IV. CONCLUSIONS

Testing chips for high power 3D ICs have been designed and fabricated. The chips contain 20W and 10W heaters with temperature sensors. Cooling blocks contain microchannel with silicon pillars to increase turbulence and enhance the cooling efficiency. Diamond layers were deposited to enhance heat transfer process. The results show that using R22 refrigerant, the temperature of a 420 W IC can be reduced to the normal operating temperature.

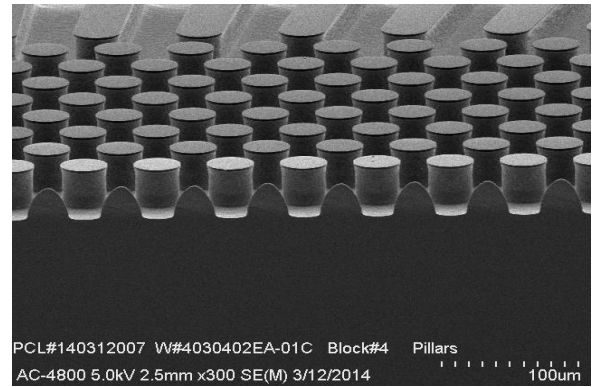

Fig. 6. SEM images of one half of cooling block, where silicon pillars of diameter 35 µm in the form of circular pillars in block #4 is shown.

TABLE II. MEASURED RESISTANCE AND TEMPERATURE

| 10 Watt heater (Center of the wafer) Wafer #4052705-002 | | | | | |
|---|---|---|---|---|---|
| Electrical Input | | Sensor Resistance (Ω) | | Temperature (˚C) | |
| Voltage (V) | Current (A) | Left | Center | Left | Center |
| 0 | 0 | 226.45 | 224.4 | 27 | 27 |
| 6 | 0.1 | 234.32 | 237.43 | 36.03 | 42.08 |
| 10 | 0.2 | 249.32 | 250.39 | 53.23 | 57.08 |
| 20 | 0.36 | 256.46 | 276.24 | 61.42 | 87 |
| 26 | 0.46 | 267.96 | 309.28 | 74.62 | 125.25 |
| 30 | 0.5 | 285.85 | 322.35 | 95.13 | 140.38 |
| 40 | 0.54 | 311.22 | 277.99 | 124.2 | 204.65 |

TABLE III. HEATERS TEMPERATURE IN (˚C) AFTER 2 PHASE COOLING

| Row/Col | 1 | 2 | 3 | 4 | 5 | 6 |
|---|---|---|---|---|---|---|
| 1 | 41 ˚C | 41 ˚C | 42 ˚C | 42 ˚C | 43 ˚C | 52 ˚C |
| 2 | 41 ˚C | 41 ˚C | 42 ˚C | 42 ˚C | 43 ˚C | 52 ˚C |
| 3 | 41 ˚C | 41 ˚C | 42 ˚C | 42 ˚C | 43 ˚C | 52 ˚C |
| 4 | 41 ˚C | 41 ˚C | 42 ˚C | 42 ˚C | 43 ˚C | 53 ˚C |
| 5 | 41 ˚C | 41 ˚C | 42 ˚C | 42 ˚C | 43 ˚C | 50 ˚C |
| 6 | 41 ˚C | 41 ˚C | 39 ˚C | 42 ˚C | 43 | 50 ˚C |

ACKNOELEDGEMENT

The authors would like to acknowledge that all members contributed equally to this study.